\documentclass[aps,prl,reprint,superscriptaddress]{revtex4-1}
\usepackage{graphicx}
\usepackage{siunitx}

\usepackage{xcolor}
\usepackage{amsmath}
\usepackage{lipsum}
\usepackage{hyperref}
\hypersetup{
	colorlinks=true,
	linkcolor=blue,
	urlcolor=blue,
	citecolor=blue,
	pdfpagemode=FullScreen,
}
\usepackage[capitalise]{cleveref}

\usepackage{enumitem}
\usepackage{float}
\makeatletter
\def\maketitle{
	\@author@finish
	\title@column\titleblock@produce
	\suppressfloats[t]}
\makeatother

\begin{document}
\title{Observation of Temperature Independent Anomalous Hall Effect in Thin Bismuth from Near Absolute Zero to 300 K Temperature}
\author{Oulin Yu}
\affiliation{Department of Physics, McGill University, Montréal, Québec, H3A 2T8, Canada}
\author{F. Boivin}
\affiliation{Department of Physics, McGill University, Montréal, Québec, H3A 2T8, Canada}
\author{A. Silberztein}
\affiliation{Department of Physics, McGill University, Montréal, Québec, H3A 2T8, Canada}
\author{G. Gervais}
\affiliation{Department of Physics, McGill University, Montréal, Québec, H3A 2T8, Canada}

\begin{abstract}
We report our discovery of a temperature independent anomalous Hall effect (AHE) from 15 mK to 300 K temperature occurring in a 68 nm thick transport device made out of  pure bismuth. This surprising behaviour is accompanied with an expected temperature dependent longitudinal resistance  consistent with  semi-metallic  bismuth, however it surprisingly showed no hint of a magnetoresistance for magnetic fields between $\pm30$ T. Even though bismuth is a diamagnetic material which  {\it a priori}  does not break time-reversal symmetry (TRS), 
our analysis of the reconstructed conductivities points towards the AHE to be of the intrinsic type,  which does not emanate from magnetic impurities. Finally, as pure bismuth has been shown numerically to host a Berry curvature at its  surface which breaks inversion symmetry, we propose it  as a possible explanation for the temperature independent AHE observed here.
\end{abstract}
\date{\today}
\pagenumbering{arabic}
\maketitle
\setlength{\parskip}{1em}

\par \textit{Introduction.---} Ever since Faraday's discovery of its diamagnetism in the mid-19\textsuperscript{th} century, elemental bulk bismuth has been extensively studied and has been a wonder material for exploring new physical properties and phenomena such as the Shubnikov-de Haas (SdH), de Haas-van Alphen and Nernst–Ettingshausen effects. More recently, there has been a revival of experimental and theoretical work on single crystalline bismuth as it has demonstrated evidence for superconductivity at temperatures below $T_c\approx0.5$ mK~\cite{Prakash_2017}, as well as edge and hinge states that are consistent with models of higher-order topology~\cite{Aggarwal_2021,Schindler_2018}. These discoveries make elemental bismuth an interesting platform for exploring exotic physical properties. For instance, a non-linear Hall effect was recently discovered in bismuth which showed tunability even at room temperature~\cite{Makushko_2024}. Additionally, being a material with high spin-orbit coupling, bismuth is known to host a buckled honeycomb lattice making it a suitable candidate to study the parity anomaly predicted by Haldane~\cite{Haldane_1988}.

\par While bismuth has been extensively studied in its bulk form, the exploration of its two-dimensional form has been limited due to fabrication challenges. Unlike graphite or other group V elements such as black phosphorus, the strong interlayer bonds make it difficult to mechanically exfoliate thin bismuth. Recent works on thin bismuth largely rely on molecular beam epitaxy (MBE)~\cite{Yang_2020,Abdelbarey_2020,Abdelbarey_2021} or more creative methods such as the use of nano-molds~\cite{Chen_2024} to press bulk bismuth down to 10 nm. These methods are not widely accessible compared to the simpler mechanical exfoliation, leaving the electronic transport properties of single crystal bismuth below 100 nm largely unexplored. In this letter, by making use of a newly developed mechanical exfoliation technique based on micro-trenches~\cite{Yu_2023}, we successfully fabricated thin bismuth devices and carried out transport measurements in a magnetic field $B$ ranging from $-30$ T to $30$ T and from near absolute zero to 300 K temperature. Strikingly, the anomalous Hall effect (AHE) discovered in our 68 nm van der Pauw (vdP) device~\cite{Yu_2024} is found to be temperature independent, while the magnetoresistance is found to be entirely featureless. Both are remarkably surprising given the wide temperature range of investigation and the large applied magnetic fields.

\begin{figure}[b!]
\centering
\includegraphics[width=0.51\textwidth]{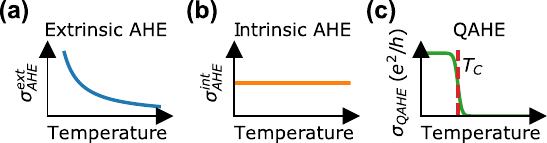}
\caption{\textbf{(a)} Example of Hall conductivity due to the extrinsic contributions. Note that the $\sigma_{\text{AHE}}^{ext}$ could be increasing or decreasing with temperature depending upon the material~\cite{Miyasato_2007}. \textbf{(b)} Hall conductivity due to the intrinsic contribution. \textbf{(c)} Below the critical temperature $T_C$, the QAHE yields exactly one conductance quantum $e^2/h$ per edge or surface state. }\label{fig:1}
\end{figure}

\par \textit{Anomalous Hall Effect.---} It is widely accepted in the literature that the total anomalous Hall conductivity (AHC) can be expressed as~\cite{Tian_2009}
\begin{equation}\sigma_{\text{AHE}}=\sigma_{sk}+\sigma_{sj}+\sigma_{int},\label{eq:sigma_AHE}\end{equation}
where $\sigma_{\text{AHE}}$ is the total AHC and $\sigma_{sk}$, $\sigma_{sj}$, $\sigma_{int}$ are the skew-scattering, side-jump, and intrinsic contributions respectively. One way to parse the mechanisms responsible for the AHE is to study how the AHC scales with the longitudinal conductivity $\sigma_{xx}(T)$ ~\cite{Nagaosa_2010,Cucler_2024,Tian_2009,Onoda_2006}. Since the skew-scattering and side-jump mechanisms both depend on impurities, the intrinsic contribution is, in principle, the only term that is temperature independent ~\cite{Shitade_2012}. A hallmark of the intrinsic contribution $\sigma_{int}$ is that it does not scale with the longitudinal conductivity $\sigma_{xx}(T)$ as described by the semi-empirical relation~\cite{Tian_2009,Nagaosa_2010}
\begin{equation}
-\sigma_{\text{AHE}}(T)=(\alpha\sigma_{xx0}^{-1}+\beta\sigma_{xx0}^{-2})\sigma_{xx}^2(T)+\sigma_{int},\label{eq:sigma_AHE2}
\end{equation}
where $\alpha$ and $\beta$ are coefficients for the skew-scattering and side-jump mechanisms, respectively, $\sigma_{xx0}$ is the zero-field residual conductivity as $T\to0$, and $\sigma_{int}$ is the intrinsic AHC that remains constant. The expected temperature dependences for the extrinsic and intrinsic dominated AHE are shown in \cref{fig:1}\textbf{(a)} and \textbf{(b)}, respectively. Depending on the material studied and the temperature range of investigation~\cite{Miyasato_2007}, the extrinsic AHE $\sigma_{\text{AHE}}^{ext}$ can increase or decrease with temperature. For comparison, the expected temperature dependence for the QAHE is also shown in \cref{fig:1}\textbf{(c)}. It is known that the intrinsic AHE is a precursor of the QAHE ~\cite{Nagaosa_2010,Weng_2015,Liu_2016} which yields an anomalous conductance of exactly $e^2/h$ per edge or surface state and is temperature independent below a critical temperature $T_C$~\cite{Chang_2023}.

\begin{figure}[!t]
\centering
\includegraphics[]{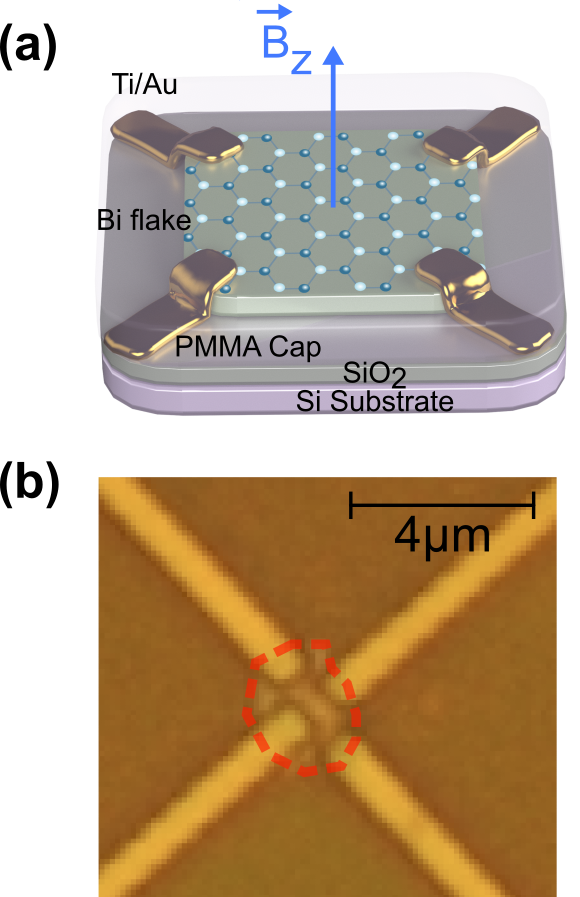}
\caption{\textbf{(a)} Schematic of the device with an arrow showing the direction of the applied magnetic field $\vec{B}_z$. \textbf{(b)} Optical microscope image of the fabricated device where the flake is circled by the red dashed line. }\label{fig:2}
\end{figure}

\begin{figure*}[!t]
\centering
\includegraphics[]{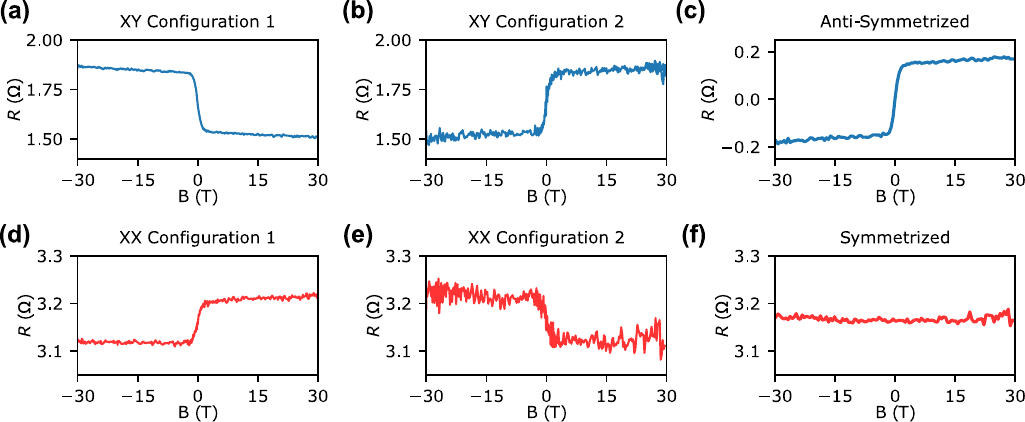}
\caption{Resistances as a function of the magnetic field at $T=1.4$ K. \textbf{(a)}, \textbf{(b)} Onsager configurations optimizing the $XY$ or Hall measurements. \textbf{(d)}, \textbf{(e)},  Onsager configurations optimizing the $XX$ or the longitudinal measurements. \textbf{(c)} Anti-symmetrized data that shows the extracted Hall resistance from the $XY$ configurations shown in \textbf{(a)}, \textbf{(b)}. \textbf{(f)} Symmetrized data that shows the extracted longitudinal resistance from the $XX$ configurations shown in \textbf{(d)}, \textbf{(e)}. Note that the noise in \textbf{(c)} and \textbf{(f)} are different due to the averaging process required for the anti-symmetrization and symmetrization.}\label{fig:3}
\end{figure*}

\par \textit{Methods.---} Details of the fabrication process can be found in the \textit{Supplemental Material} (SM) and in Yu \textit{et al.}~\cite{Yu_2023} which also includes the AHE observations in two other bismuth devices in a comb geometry with thicknesses of 29 nm and 69 nm. The illustration and the optical image of the 68 nm vdP device is shown in \cref{fig:2}\textbf{(a)} and \textbf{(b)}, respectively. Transport measurements (details found in the SM) are performed in a perpendicular magnetic field with field strengths up to $\pm30$ T and temperatures ranging from 1.4 K to 300 K in a resistive magnet. The lowest temperature data at $T=15$ mK was measured between $\pm 9$ T in a dilution refrigerator~\cite{Yu_2024} showing  identical behaviors~\cite{Yu_2024} , though for the remainder of the manuscript, we will focus only on the 1.4 K to 300 K data as they were taken during a single cooldown. 

\par \textit{Onsager Symmetrization.---}As shown in \cref{fig:2}\textbf{(b)}, the Ohmic contacts deposited on the bismuth flake are located close to one another due to the small size of the flake. An ideal vdP geometry would have the contact electrodes located at each corner of a square sample, as depicted in the insets of \cref{fig:3}, and this is not the case here. Consequently, in our measurements, the Hall ($R_{xy}$) and longitudinal ($R_{xx}$) resistances are inevitably mixed and are measured simultaneously regardless of the probe configuration. To remedy this, we made use of Onsager's reciprocity theorem ~\cite{Sample_1987} to reconstruct the Hall and longitudinal resistances, as discussed in detail in the SM. 

\par The resistances as a function of the magnetic field were measured in two different Onsager pairs: one that maximizes the Hall signal ($XY$ configurations 1 and 2) as shown in \cref{fig:3}\textbf{(a)} and \textbf{(b)}, and one that maximizes the longitudinal signal ($XX$ configurations 1 and 2) as shown in \cref{fig:3}\textbf{(d)} and \textbf{(e)}. The $XY$ Onsager pair's data was then anti-symmetrized to obtain the Hall resistance, shown in \cref{fig:3}\textbf{(c)}.  Similarly, the $XX$ Onsager pair's data was symmetrized to extract the longitudinal resistance, as shown in \cref{fig:3}\textbf{(f)}.

\par \textit{Hall and Longitudinal Resistances.---}As previously reported~\cite{Yu_2024} and as shown in \cref{fig:3}\textbf{(c)}, the AHE is unambiguously observed in our elemental bismuth device. Hints of the AHE and/or anomaly in electronic magnetotransport occurring in bismuth fibres were reported  in the 1940's by Conn and Donovan~\cite{CONN_1948,Donovan_1950}, as well as recently in its bulk form by Camargo \textit{et al.}~\cite{Camargo_2021} who eliminated the possibility of magnetic contaminants and speculated that the AHE arises from the surface. In bismuth thin films, the AHE has been demonstrated by Hirai \textit{et al.}  by opening a gap in bismuth's band structure with intense circularly polarized light~\cite{Hirai_2023}. Another more recent instance of the AHE in bismuth is  in the work by Abdelbarey \textit{et al.}~\cite{Abdelbarey_2020} where the authors  clearly observed  a non-linearity of the Hall signal near zero-field.\\

\par Interestingly, in Partin \textit{et al.}'s work on MBE-grown bismuth films~\cite{Partin_1988}, although it was overlooked by the authors, their 100 nm film displayed a Hall response and a completely featureless longitudinal response up to 17 T that agreed with what we have found in our 68 nm device between a  magnetic field of $\pm30$ T (see \cref{fig:3}\textbf{(f)}). While their film were unlikely to be protected by an encapsulating layer, all of our devices were capped by a PMMA layer in order to avoid oxidization and aging. These converging observations indicate that the AHE in elemental bismuth is genuine, yet as of today its exact mechanism remains unclear.

\par The complete absence of magnetoresistance and SdH oscillations is particularly puzzling since diamagnetic and semi-metallic materials typically have a magnetoresistance that scales with $B^2$ ~\cite{Cabrera_Baez_2023,Juraszek_2019,Leahy_2018,Liang_2014,Ali_2014,Shekhar_2015,Kumar_2017,Tafti_2015}. Meanwhile, SdH oscillations were observed in bulk bismuth for magnetic fields as low as 1 T~\cite{Lerner_1962}, and in a more recent work on growing $10$ to $20$ nm thin bismuth in nano-molds at magnetic  fields as low as 5 T~\cite{Chen_2024}. As pointed out in the SM and in our previous report at lower magnetic field~\cite{Yu_2024}, our observation of a featureless longitudinal magnetoresistance for magnetic fields between $\pm 30$~T cannot be explained by a multi-carrier model and rather points toward an unknown mechanism being at play.\\

\begin{figure*}[!ht]
\centering
\includegraphics[scale=1]{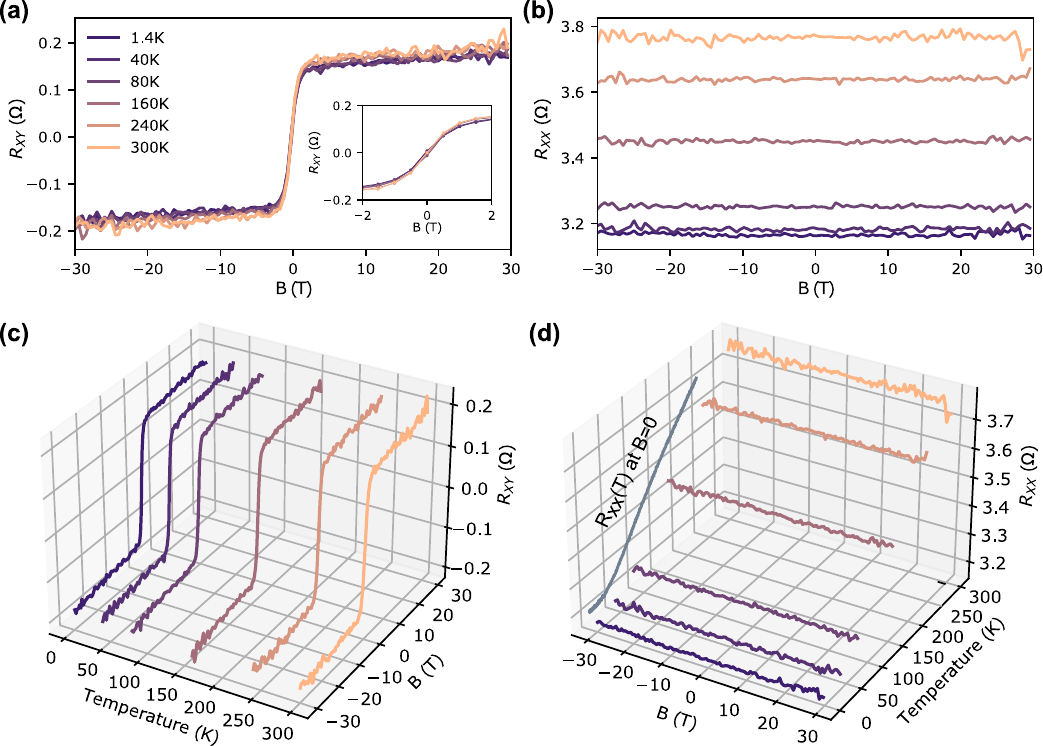}
\caption{\textbf{(a)} Hall and \textbf{(b)} longitudinal resistances as a function of the magnetic field at various temperatures ranging from 1.4 K to 300 K. The inset in \textbf{(a)} is the zoom-in for the data between $\pm 2$ T. \textbf{(c)}, \textbf{(d)}, 3D plots of the same curves in \textbf{(a)} and \textbf{(b)} respectively. The longitudinal resistance $R_{xx}(T)$ for the $XX$ configuration 1 is shown in the background of \textbf{(d)} for comparison. }\label{fig:4} 
\end{figure*}

\par \textit{Temperature Independence of the AHE in Thin Bismuth.---}The main result of this work is shown in \cref{fig:4}. The resistances versus magnetic field for all vdP configurations shown in \cref{fig:3} were measured at temperatures of 1.4 K, 40 K, 80 K, 160 K, 240 K and 300 K during a single cooldown. The Hall resistances $R_{xy}$ extracted via anti-symmetrization are shown in the main panel of \cref{fig:4}\textbf{(a)} where the inset shows a zoom-in  at magnetic fields between $\pm 2$ T. The same $R_{xy}$ data is shown in three dimensions in \cref{fig:4}\textbf{(c)}. Similarly, the longitudinal resistance extracted from the $XX$ Onsager pair is plotted in \cref{fig:4}\textbf{(b,d)}. In \cref{fig:4}\textbf{(d)}, the zero field ($B=0$ T) temperature dependence of $R_{xx}(T)$ for the $XX$ configuration 1 between 3 K and 260 K is shown in gray in the background.  This increasing resistance with increasing temperature is consistent with the semi-metallic nature of bulk and thin bismuth ~\cite{Partin_1988,Hoffman_1971,Kukkonen_1977}. 

\par Most strikingly, our data shown in \cref{fig:4}\textbf{(a,c)} demonstrates the observed AHE to be, within noise, entirely independent of temperature. The negligible temperature dependence is particularly notable in the low field region of $|B|< 2$ T where the $R_{xy}(B)$ data completely overlap for all temperatures investigated. This was further verified down to 15 mK in a dilution refrigerator during a separate cooldown \cite{Yu_2024}. In the high field regions where the AHE is saturated, the anomalous Hall resistance $R_{\text{AHE}}$ can be extracted via the zero field intercept of a linear fit (see SM). At first sight, this lack of temperature dependence may seem surprising. However, an important feature of the intrinsic AHE is that it is temperature independent~\cite{Nagaosa_2010,Tian_2009}, as illustrated in \cref{fig:1}\textbf{(b)}. This is further expected in our case since the estimated longitudinal conductivity of our device is in the good-metal regime $\sigma_{xx}\sim10^4-10^6$ \si{\per\ohm\per\centi\meter}, which is typical for materials exhibiting an AHE that is intrinsic in nature~\cite{Nagaosa_2010}. Furthermore, our analysis on the relation found in \cref{eq:sigma_AHE2}  (see SM) shows the proper linear scaling between $-\sigma_{\text{AHE}}$ and $\sigma_{xx}^2$, further pointing the observed AHE towards the intrinsic type. \\

\par The underlying mechanism of the AHE in bismuth is,  in our view, unlikely due to magnetism of any kind given the lack of temperature dependence. Furthermore, as discussed in more detail in the SM, elemental bismuth is known to be a diamagnetic material which {\it a priori} does not break TRS and has a relatively weak temperature dependent susceptibility that varies by a factor of two from 540 K to low temperatures~\cite{Otake_1980}.  In contrast,  an intrinsic AHE can originate from a non-zero Berry curvature~\cite{Nagaosa_2010,Cucler_2024} with the necessary condition of breaking either TRS or inversion symmetry (IS)~\cite{Xiao_2010}. In bismuth, IS is preserved in the bulk but broken at the surface~\cite{Hofmann}, thus a non-zero Berry curvature could be present at the surface without breaking TRS. Recent first principle calculations by Wawrzik \textit{et al.}~\cite{Wawrzik_2023} confirmed this by showing that while bismuth has a zero Berry curvature in the bulk, its surface nevertheless hosts a non-zero Berry curvature. This serves as the basis for understanding the room temperature non-linear Hall effect observed very recently by Makushko et al.~\cite{Makushko_2024}. We propose that bismuth's non-zero Berry curvature at its surface could explain the temperature independent AHE observed in our work. Future theoretical and experimental work is nevertheless  required to  pinpoint and gain knowledge on the exact mechanism being here at play.\\

\par \textit{Conclusion.---}In this work, we measured the resistances of a 68 nm bismuth transport device in different vdP configurations for magnetic fields between  $\pm30$ T and temperatures between 15 mK and 300 K. The extracted longitudinal resistance was observed to be completely featureless for all investigated temperatures. The extracted Hall resistance clearly demonstrates an anomaly consistent with the anomalous Hall effect, and was observed to be entirely independent of temperature. Most, if not all mechanisms, except for the Berry curvature responsible for the intrinsic AHE, are temperature dependent in the range of investigation. We are unaware of any model that could explain simultaneously the observed Hall and longitudinal signals, nor any other material that exhibit an electronic Hall transport that is so insensitive to temperature for such a wide range. This makes bismuth an intriguing material that is still not well understood. Looking forward, given that intrinsic AHE is the precursor of QAHE~\cite{Weng_2015,Liu_2016}, we speculate whether bismuth could be a suitable platform to explore the QAHE given its buckled honeycomb crystal structure and high spin-orbit coupling, both being important ingredients for the parity anomaly predicted by Haldane~\cite{Haldane_1988}.  If this is the case for bismuth, then this also raises the fascinating question of whether the QAHE could be observable at much higher temperatures than it currently has been.

\par \textit{Acknowledgment.---}This work has been supported by  the Natural Sciences and Engineering Research Council of Canada, the New Frontiers in Research Fund program, the Canadian Foundation for Innovation, Montréal-based CXC and the Fonds de Recherche du Québec Nature et Technologies. Sample fabrication was carried out at the McGill Nanotools Microfabrication facility and at the Laboratoire de Microfabrication at Polytechnique Montréal. A portion of this work was performed at the National High Magnetic Field Laboratory, which is supported by National Science Foundation Cooperative Agreement No. DMR-2128556* and the State of Florida. We are grateful to M.-O. Goerbig, T. Szkopek, and Z. Berkson-Korenberg for the extremely helpful discussions and comments, as well as for the reading of the manuscript. Finally, we would like to thank R. Talbot, R. Gagnon, J. Smeros and S. Chakara for technical assistance.

\clearpage
\newpage\pagebreak
\title{Supplemental Material: Anomalous Hall Effect in Thin Bismuth}
\maketitle
\onecolumngrid
\setcounter{figure}{0}
\renewcommand\thefigure{S\arabic{figure}}

\section*{Device Fabrication and Characterization}
\begin{figure}[h!]
	\centering
	\includegraphics[width=0.5\textwidth]{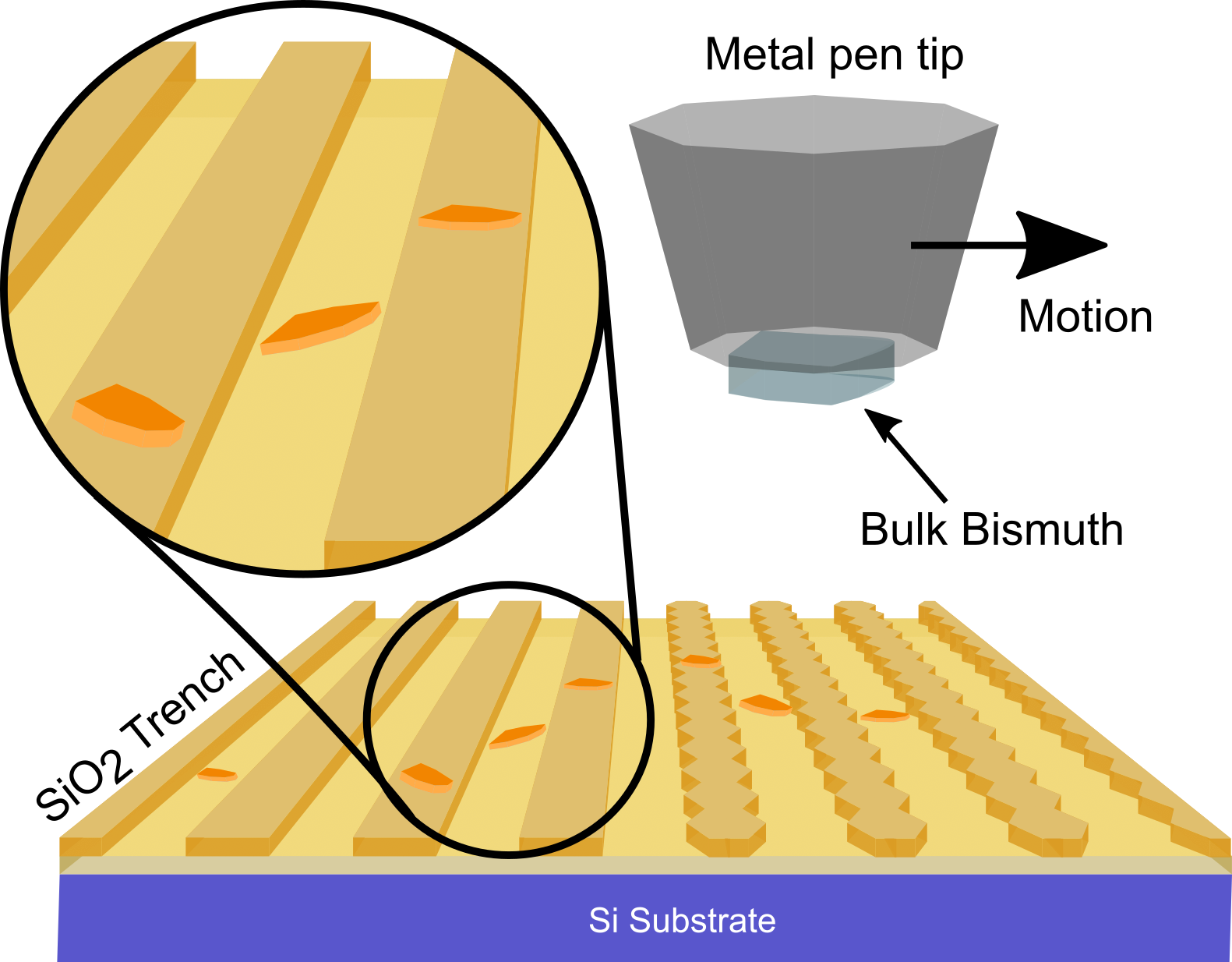}
	\caption{Schematic of the mechanical exfoliation process that uses a micro-trench structure. A bulk single crystal bismuth is glued to a pen which is then used to grate against a SiO$_2$ surface with micro-trenches in order to exfoliate thin bismuth flakes.}\label{fig:S1}
\end{figure}
\par Thin bismuth flakes were exfoliated with our newly-developed mechanical exfoliation technique that utilizes micro-trench structures \cite{methodpaper_SM}. To do so, a bulk bismuth single crystalline sample with 99.9\% purity was attached to a metal pen, as shown in Fig. S1. This pen was then used to grate the bulk bismuth against a SiO$_2$ surface patterned with micro-trenches designed to facilitate the breakage of interlayer bonds in bismuth as depicted in Fig. S1. With this method, it is possible to produce thin flakes with thicknesses down to the 2 nm range. Importantly, the entire exfoliation process was performed in a nitrogen-only glovebox with O$_2$ and water concentrations both less than 1 ppm, and as such oxidation of the exfoliated bismuth flakes was avoided. Additionally, a polymethyl methacrylate (PMMA) protection capping layer was deposited after the entire fabrication process in order to further protect the device from oxidation. The bismuth nature of the fabricated devices was confirmed with Raman spectroscopy and benchmarked with the parent bulk crystal as well as bismuth oxide. In particular, the Raman spectroscopy showed no sign of bismuth oxide or aging effects for a period of more than two years since the devices' fabrication. Further details of our fabrication technique and pertaining to the Raman characterization can be found in Ref. \cite{methodpaper_SM,devicepaper_SM}. As a final step, the selected flakes were patterned with 15nm/100nm Ti/Au contacts and then capped again with a PMMA protection layer.

\par The bismuth flakes were characterized with atomic force microscopy (AFM) in order to identify the most promising flakes for electronic transport devices. Figures S2\textbf{(a)},\textbf{(b)} show the optical microscopy image and the AFM measurement of the bismuth device presented in the main manuscript. In Fig. S2\textbf{(b)}, a height profile is selected across the bismuth flake which is shown in Fig. S2\textbf{(c)}. Note that the height profile shows a non-negligible height variation, and by comparing with the active channel in the final device (Fig. 2\textbf{(b)} of the manuscript), the average height is calculated to be $68\pm47$ nm. The large uncertainty is due to the wedge shape of the sample shown in Fig. S2\textbf{(c)}. Despite this large uncertainty, we do not expect this height variation to impact the transport measurements in a significant way, because we also observed the anomalous Hall effect in other devices fabricated using the same technique, yet with different thickness ($29\pm19$ nm and $69\pm20$ nm)~\cite{devicepaper_SM}.  
\begin{figure}[h!]
	\centering
	\includegraphics[]{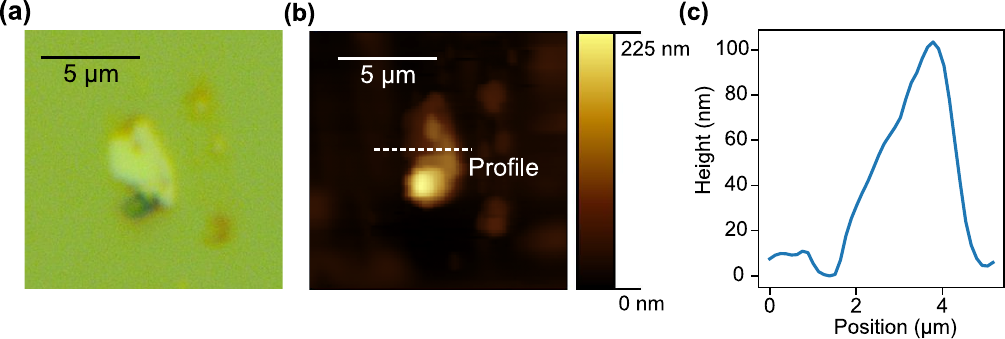}
	\caption{\textbf{(a)} Original optical image of the device flake. \textbf{(b)} The corresponding AFM scan of the same flake. \textbf{(c)} The height profile as selected in the AFM scan.}\label{fig:S2}
\end{figure}

\section*{Extraction of the Longitudinal and Hall Resistances}
\par To extract the Hall and longitudinal resistances that were mixed due to the small size of the sample and contact misalignments, we use a symmetrization based on Onsager's reciprocal relations~\cite{Sample_1987_SM}. Specifically, Onsager symmetrization~\cite{Sample_1987_SM} states that in a linear system, the resistance tensor
$$R =\left(\begin{matrix}
R_{xx} & R_{xy}\\
-R_{xy} & R_{xx}\\
\end{matrix}\right)$$
is transposed upon reversal of the current and voltage probes. Therefore, measuring the same contacts twice while reversing the current and voltage probes allows the decoupling of the longitudinal ($R_{xx}$) and the Hall ($R_{xy}$) resistances by symmetrizing and anti-symmetrizing the raw data. Specifically, this is achieved via the following formulas:
\begin{align*}
R_{xx}&=\frac{R+R'}{2},\\
\text{ and }\qquad R_{xy}&=\frac{R-R'}{2},
\end{align*}
where $R$ and $R'$ form an Onsager reciprocal pair whose voltage and current leads are reversed, \textit{i.e.} $V_+\longleftrightarrow I_+$ and $V_-\longleftrightarrow I_-$.
\par Note that for the symmetrization and anti-symmetrization, wed create $N=120$ bins between $B=-30$ T and $B=30$ T. $R$ and $R'$ are averaged for these bins, then plugged into the above equations to calculate $R_{xx}$ and $R_{xy}$.

\begin{figure}[h!]
	\centering
	\includegraphics[]{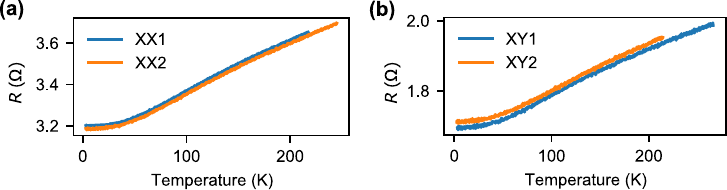}
	\caption{Temperature dependence of the resistance for the \textbf{(a)} XX Onsager pairs and \textbf{(b)} XY Onsager pairs.}\label{fig:S3}
\end{figure}

\section*{Temperature Dependence}
\par Recall that Fig. 4\textbf{(d)} in the manuscript shows the dependence of longitudinal resistances on the magnetic field at various temperatures, and in the background, a raw XX four-point resistance's temperature dependence is shown for comparison. For completeness, here we show in Fig. S3 the full temperature dependences for all XX and XY Onsager pairs that were used for the symmetrization and anti-symmetrization procedures to obtain the XX and XY electronic transport components (see Fig. 3\textbf{(c, f)} of the main manuscript).

\par In Fig. 3 of the manuscript, we present the raw resistances versus the magnetic field for XX and XY Onsager pairs at $T=1.4$ K. For completeness, Fig. S4 shows the raw resistances versus the magnetic field for these Onsager pairs at temperatures 1.4 K, 40 K, 80 K, 160 K, 240 K and 300 K. 
\begin{figure}[h!]
	\centering
	\includegraphics[width=0.8\textwidth]{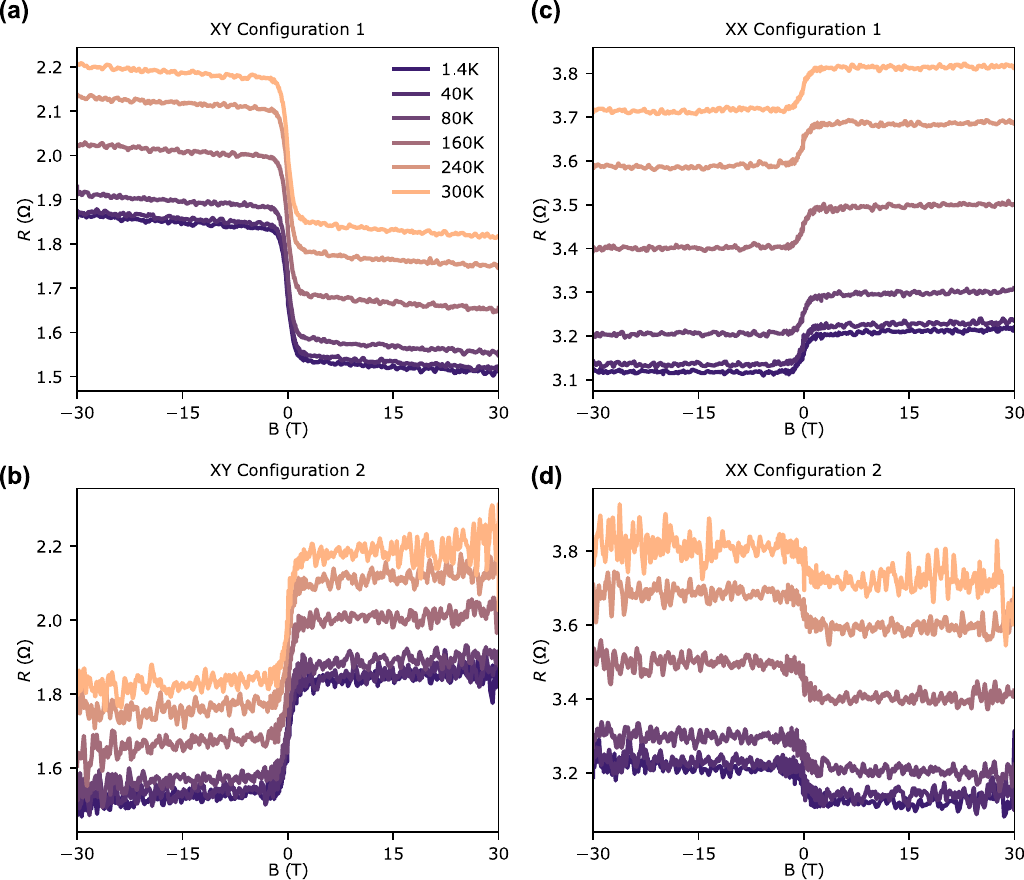}
	\caption{Raw resistances versus the magnetic field for \textbf{(a, b)} the XY Onsager pairs and \textbf{(c, d)} the XX Onsager pairs. }\label{fig:S4}
\end{figure}
\section*{Multi-Carrier Model for the Hall and Longitudinal Resistances}

\par While a single valley system can potentially explain the field-independent $\rho_{xx}$~\cite{Zhu_2018_SM} observed in our work, this would lead to a Hall coefficient that is unreasonably small (for example $0.00106(13)$ \si{\ohm\per\tesla} at 1.4 K). In addition, bulk bismuth is also known to host both electron and hole pockets~\cite{Jain_1968_SM,Jain_1962_SM}, and as a result a multi-carrier model can only exhibit a featureless $\rho_{xx}$ at high magnetic fields when the carrier concentrations are different, \textit{i.e.} when $n\ne p$~\cite{Zhu_2018_SM}. This occurs when the field strength exceeds a characteristic field 
\begin{equation}B^\star =\dfrac{n \nu + p \mu}{|n-p|\mu \nu},\label{eq:bstar_SM}\end{equation}
where $n$, $p$ are the electron and hole concentrations and $\nu$, $\mu$ are the electron and hole mobilities. When $|B|\le B^\star$, a $B^2$ dependence~\cite{Zhu_2018_SM} is nevertheless expected, and this is clearly not observed in the data presented in Fig. 3\textbf{(f)} of the main text. Furthermore, we note that it is not possible to fit the parameters mentioned above, because of the complete lack of explicit dependence on the magnetic field for $\rho_{xx}$, and therefore this is uncharacteristic of two carrier types that would be contributing to the electronic transport. We therefore believe that a more elaborate model is required to understand and explain both the longitudinal resistivity that shows no sign of magnetoresistance as well as the AHE observed in our work.

\section*{Effect of Diamagnetism of Bismuth}
\par Since bismuth is diamagnetic, it induces an opposing magnetization $M$ proportional to the external applied field $\mu_0H$ given by $M=\chi H$ where $\chi$ is the magnetic susceptibility. The effective field $B$ is then
\begin{equation*}
B = \mu_0 H  + \mu_0M = \mu_0H(1 + \chi).
\end{equation*}
In bismuth, as seen in Fig. S5, $\chi$ depends on temperature and crystal direction. For temperatures below 300 K, 
\begin{equation*}
-2.4 \times 10^{-4} < \chi < -1.3 \times 10^{-4},
\end{equation*}
which gives,
\begin{equation*}
0.99976 < 1 + \chi <  0.99987.
\end{equation*}
This small correction factor means that any modification taking into account the difference between the external and effective field falls within the noise of our signal.

\begin{figure}[h!]
	\centering
	\includegraphics[]{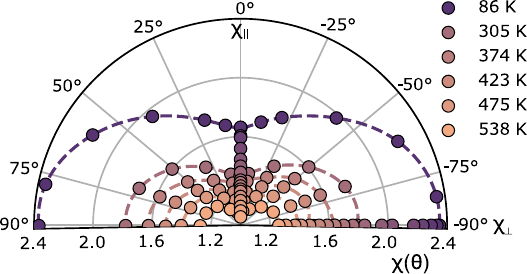}
	\caption{The angular and temperature dependences of the diamagnetism of bismuth. Data digitized from Otake \textit{et al.}~\cite{Otake_1980_SM}.}\label{fig:polar}
\end{figure}

\section*{Extraction of the Hall coefficient and Anomalous Hall Resistance}
\par The anomalous Hall resistance is defined as the zero-field intercept of $R_{xy}$. To obtain this value, we fit the saturated part of $R_{xy}$ ($|B|>2$ T) and average the absolute values of the intercepts as shown in Fig. S5 for $T=1.4$ K, \textit{i.e.}
$$R_{\text{AHE}}=\dfrac{0.147(1)+0.146(1)}{2}=0.146(2) \text{ }\si{\ohm},$$
where the errors are quoted from the linear regression fits and shown in parentheses. From the slopes fitted in Fig. S6, we average the two slopes to find the Hall coefficient provided the main text, \textit{i.e.}
$$R_H=\dfrac{0.00099(6)+0.00114(7)}{2}=0.00106(13)\text{ } \si{\ohm\per\tesla}.$$
The same process was then repeated at all other temperatures chosen to perform the electronic transport measurements.

\begin{figure}[h!]
	\centering
	\includegraphics[]{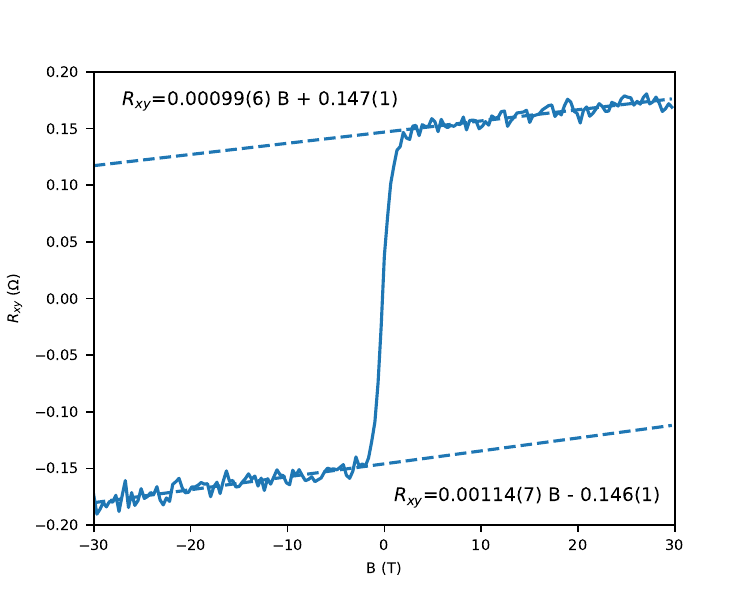}
	\caption{Extraction of the Hall coefficient and $R_{\text{AHE}}$ from $R_{xy}(B)$. The linear fits are restricted to the range $|B|>2$ T. The fitted results are shown in the graph with the errors from the linear regression shown in parenthesis.}\label{fig:S5}
\end{figure}

\section*{Scaling of the Anomalous Hall Effect}

\begin{figure}[h!]
	\centering
	\includegraphics[]{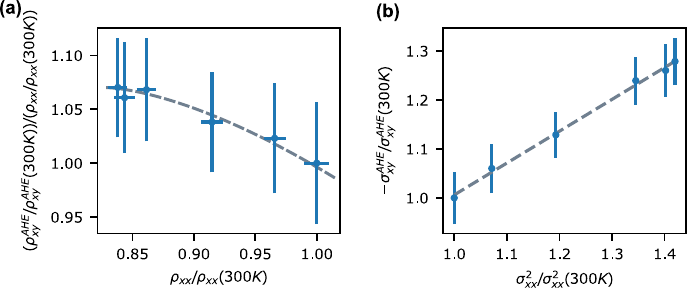}
	\caption{\textbf{(a)} The departure from a linear $\rho_{\text{AHE}}/\rho_{xx}$ versus $\rho_{xx}$ indicates that \cref{eq:rho_sk1_SM} most likely does not apply in our case (see text below). The dashed gray line is a guide for the eye. \textbf{(b)} The linearity (with the linear fit shown by the gray line) of $\sigma_{xy}^{AHE}$ vs $\sigma_{xx}^2$ confirms the validity of \cref{eq:sigma_AHE2_SM}.}\label{fig:S6}
\end{figure}

\par Given the non-trivial geometry in our transport device, it is challenging to extract confidently the geometric factor required to convert resistance to resistivity and conductance to conductivity. However, it can be observed in Fig. 3\textbf{(c, f)} of the manuscript that $\rho_{xx}\gg \rho_{xy}$, and in this regime, the intrinsic anomalous Hall resistivity $\rho^{\text{int}}_{\text{AHE}}$ is expected to scale with $\rho_{xx}^2$~\cite{Nagaosa_2010_SM,Tian_2009_SM}, \textit{i.e.} \begin{equation}\rho_{\text{AHE}}=a_{sk}\rho_{xx}+b\rho_{xx}^2,\label{eq:rho_sk1_SM}\end{equation}
where the first term provides the skew-scattering contribution and the second term is the intrinsic contribution to the AHE. However, in some situations, the literature states that phonons can be neglected and in such case, the AHE conductivity can be written as~\cite{Tian_2009_SM}
\begin{equation}
-\sigma_{\text{AHE}}=(\alpha\sigma_{xx0}^{-1}+\beta\sigma_{xx0}^{-2})\sigma_{xx}^2+\sigma_{int},\label{eq:sigma_AHE2_SM}
\end{equation}
where $\alpha$ and $\beta$ are coefficients for the skew-scattering and side-jump mechanisms respectively, $\sigma_{xx0}$ is the residual conductivity and $\sigma_{int}$ is the intrinsic anomalous Hall conductivity. In the spirit of not knowing the geometric factor for converting resistance to resistivity and then to conductivity, we can circumvent the verification of \cref{eq:rho_sk1_SM} and \cref{eq:sigma_AHE2_SM} in their normalized forms by dividing the 300 K data, as is shown in Fig. S7. In particular, the normalized $\rho_{\text{AHE}}/\rho_{xx}$ versus $\rho_{xx}$ is shown in Fig. S7\textbf{(a)}, and a clear deviation from linearity is observed with this formalism. Moreover, the fact that $\rho_{\text{AHE}}/\rho_{xx}$ decreases with increasing $\rho_{xx}$ implies that \cref{eq:rho_sk1_SM} does not model well our data. In contrast, Fig. S7\textbf{(b)} shows that the normalized $\sigma_{xy}^{{AHE}}$ scales linearly with  the normalized $\sigma_{xx}^2$, indicating that \cref{eq:sigma_AHE2_SM} is a better model for the AHE observed in bismuth. Importantly, this shows that phonons induce a negligible effect on the skew-scattering term. In future works, when the resistivities will be confidently extracted, this model will allow us to quantify the intrinsic contribution of the AHE, and as such, the AHE observed in bismuth most likely emanates from an intrinsic mechanism related to a Berry curvature.

\section*{Temperature Dependence of the Anomalous Hall Resistivity}
\begin{figure}[ht!]
	\centering
	\includegraphics[]{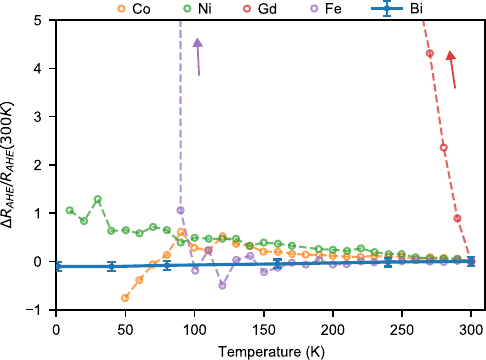}
	\caption{The change in anomalous Hall resistance in our bismuth device compared with other elemental films also in the good-metal regime~\cite{Miyasato_2007_SM}. The uncertainty in our data is due to a uncertainty in the alignment of the magnetic field and the sample's plane. Within the observed noise and errors, the data shows negligible temperature dependence in the AHE from 1.4 K to 300 K.}\label{fig:5}
\end{figure}
\par In Fig. S8, we normalize the change in $R_{AHE}$ defined by $$\frac{\Delta R_{AHE}}{R_{AHE}(300 \text{ K})} \equiv \frac{R_{AHE}(T)-R_{AHE}(300 \text{ K})}{R_{AHE}(300 \text{ K})},$$ and plot it versus the temperature to compare to other elemental materials also in the good-metal regime~\cite{Miyasato_2007_SM}. Note that the $\rho_{AHE}$ of these materials was reconstructed from the tensor relation $\hat{\rho}=\hat{\sigma}^{-1}$. In our experiment, the sample was mounted on a rotating probe that is prone to a slight uncertainty in the angle which was estimated to be $\Delta \theta\approx 5^\circ$. From the angular dependence shown in the section below, the error bars from this angular uncertainty were then estimated for the bismuth data shown in Fig. S8. We emphasize that our $R_{AHE}$ is extremely insensitive to temperature variations even compared to the materials shown in Fig. S8 which are believed to exhibit the intrinsic type of the AHE. Such temperature independence is also reminiscent of the QAHE whose resistance is quantized to $h/e^2$ at temperatures below a critical temperature $T_C$. Curiously, bismuth has recently been reported to be a higher order topological insulator~\cite{HOTI1_SM,HOTI2_SM} and a Weyl semi-metal~\cite{BiWeyl_SM}, and we note that both are systems that can exhibit the AHE and the QAHE~\cite{WeylAHE1_SM,WeylAHE2_SM}.

\section*{Resistance Uncertainty due To Angular Alignment}
\par The angle of the magnetic field with respect to the device was varied between $0^\circ$ and $90^\circ$ by rotating the chip carrier on the probe of the variable temperature insert (VTI) (see Fig. 2\textbf{(d)} of the main manuscript) while keeping the direction of the magnetic field fixed. We define $0^\circ$ when the magnetic field is entirely perpendicular to the device that is in the (111) direction of the thin bismuth. As the angle $\theta$ is varied, upon reaching $90^\circ$, the magnetic field is entirely in the plane of the Si/SiO$_2$ substrate and hence our device. A schematic of the sample rotation with respect to the magnetic field is shown in the inset of Fig. S9. 

\begin{figure}[h!]
	\centering
	\includegraphics[]{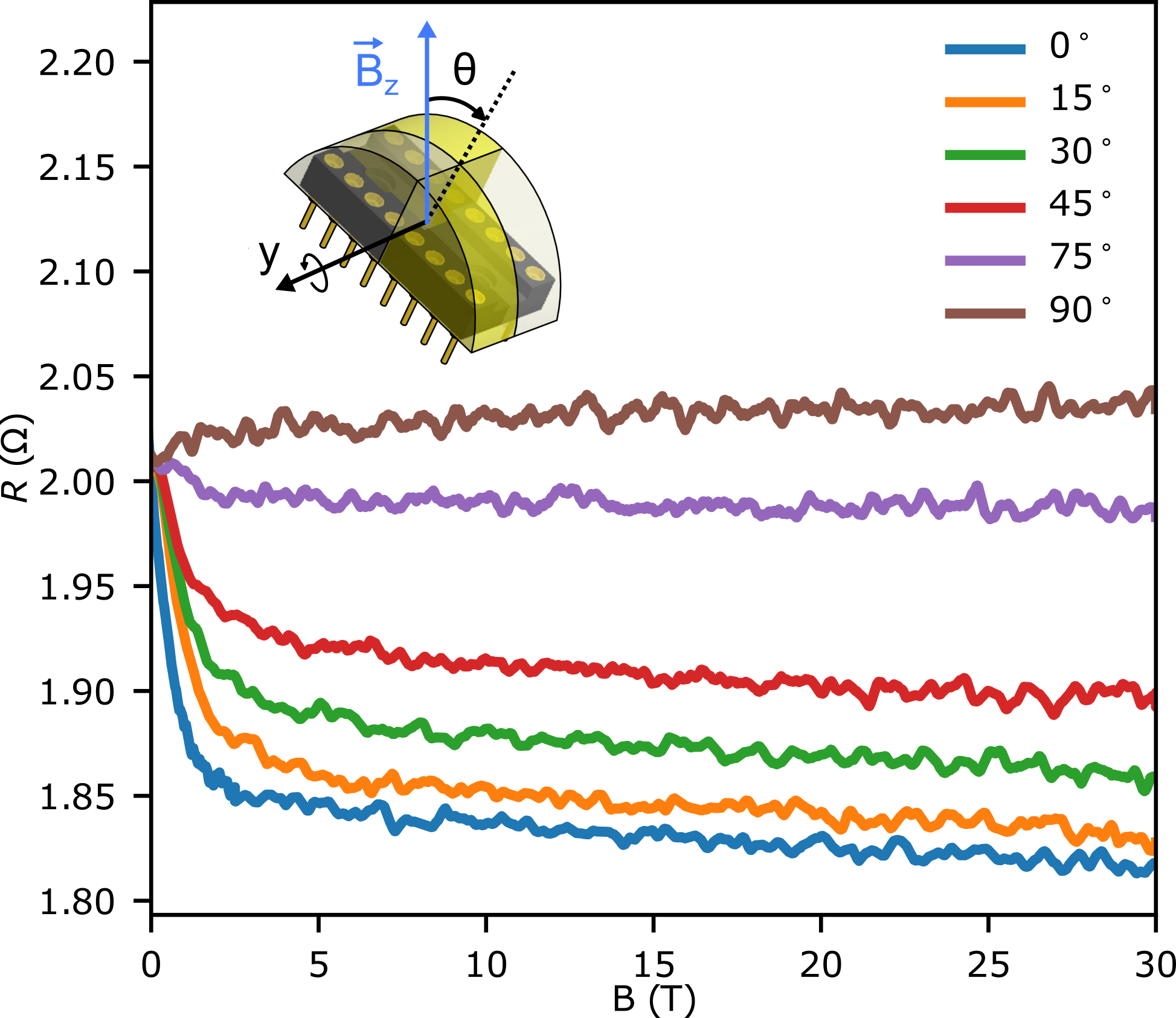}
	\caption{Resistance versus the magnetic field for the XY1 configuration at 300 K. This angular dependence is used to estimate the error bars due to angle misalignment in Fig. S8.}\label{fig:S7}
\end{figure}

\par In the XY configuration 1, the Hall signal is measured at 300 K between 0 T to 30 T and for angles $0^\circ\le\theta\le 90^\circ$. The obtained curves are plotted in Fig. S9. Note that due to the larger noise present at 300 K of the other XY configuration, we show the raw data that is not anti-symmetrized. However, the longitudinal resistance $R_{xx}$ is field-independent (see Fig. 4 of the main manuscript), and the XY configuration 1 alone captures the essence of the Hall trend despite the mixing between $R_{xy}$ and $R_{xx}$. 

\par Finally, since the error bars in Fig. S8 are dominated by the uncertainty in the angle, we can use Fig. S9 to estimate the uncertainty in $R_{AHE}$. Focusing on $\theta=0^{\circ}$ and $\theta=15^\circ$, we estimate the change in the anomalous Hall resistance to be $0.001$ \si{\ohm\per\deg}. This was then used to calculate the error bars in Fig. S8.

\end{document}